\documentstyle[aps,epsf,psfig,floats,pre,12pt,tighten]{revtex}

\begin{document}
\title{Twist and writhe dynamics of stiff filaments} \author{A.C.
  Maggs} \address{ESPCI, 10 rue Vauquelin, 75231 Paris Cedex 05,
  France.} \address{\begin{minipage}{5.55in}
\begin{abstract}
  This letter considers the dynamics of a stiff filament, in
  particular the coupling of twist and bend via writhe. The time
  dependence of the writhe of a filament is $W_r^2\sim L t^{1/4}$ for
  a linear filament and $W_r^2\sim t^{1/2} / L $ for a curved
  filament.  Simulations are used to study the relative importance of
  crankshaft motion and tube like motion in twist dynamics.  Fuller's
  theorem, and its relation with
  the Berry phase, is reconsidered for open filaments.\\
\end{abstract}
\end{minipage}\vspace*{0.1cm} 
} \maketitle

\def\s2{{\mathcal S}_2} \def\ts{{\bf t}(s)}

DNA and other stiff polymer systems such as actin filaments are of
interest both for their intrinsic biological importance, but also due
to the fact that micro-manipulation techniques give a very detailed
vision of their properties, allowing one to study fundamental
processes in polymer statics and dynamics.  Recent theoretical and
experimental work on DNA force-extension curves has shown the
importance of the static coupling between bend and twist in stiff
polymers \cite{marcandnelson,bensimon}. Non-trivial twist-writhe
correlations in flexible polymers have been recently found in a simple
dynamic lattice model in which linking number is conserved
\cite{tten}.  Here I examine the {\it dynamics} of twist and writhe
fluctuations in order to clarify the driving forces and dissipative
processes important in the motion of filaments.  I use a combination of
scaling arguments and simulations; the numerical studies are
complementary to recent work \cite{wiggins} which considers the zero
temperature motion of stiff filaments via integration of the continuum
differential equations. Here I study Brownian dynamics of stiff
filaments and introduce a generalized bead-spring model suitable for
studying twist and writhe dynamics of polymers.

There are two physical processes which can lead to the rotation of the
end of a filament about its local tangent. The two mechanisms of
spinning are {\it twist} and {\it writhe}. Twist corresponds to
excitation of internal torsional degrees of freedom of the filament.
Writhing is due to the three dimensional geometry of bending of the
filament in space.  We shall reserve the word ``spinning'' for this
process of end rotation and denote the spinning angle by $\Psi$.  The
two contributions are additive so that we can write $\Psi=\phi +W_r$,
with $\phi$ the contribution from internal twisting and $W_r$ the
contribution from the geometry of the path in space.

Excitations of internal twist degrees of freedom are well understood
\cite{zimm} and have been observed in DNA and actin via depolarized
light scattering experiments.  The writhing of a filament is best
expressed with the help of Fuller's theorem, \cite{fuller}.  Consider
a filament (with boundary conditions such that the two end tangents
are maintained parallel) which is bent into a non planar curve.  The
tangent to the filament {\bf t}(s) sweeps out a curve on the unit
sphere, $\s2$ as a function of the internal coordinate $s$.  It can be
shown, \cite{fuller}, that the contribution to the spinning due to the
writhe is just the area, $A_{\Omega}$, enclosed by ${\bf t}(s)$ on
$\s2$.

Let us assume that we are working with uniform filaments with circular
cross sections then the two process are controlled by two independent
elastic constants, the torsional stiffness, $K$ and the bending
stiffness $\kappa$. If we work in units such that $k_B T =1$ these two
elastic constants have the dimensions of lengths, and are indeed the
persistence lengths for static torsional and bending correlations of
the filament. The dynamics obeyed by the torsional and bending modes
are, however, different which leads to distinct contributions to
$\Psi$ as we shall now demonstrate.

The torsional motion of a straight filament obeys a Langevin equation
\begin{equation}
a^2{\partial  \phi \over \partial t} = K {\partial^2 \phi \over \partial s^2} + \xi (t,s) 
\label {diffusion}
\end{equation} 
where the angle $\phi(s,t)$ is the local rotation of the filament in
the laboratory frame, $\xi$ is the thermal noise, $a$ is comparable to
the radius of the filament, (we neglect all pre-factors of order unity
and use units in which $\eta = k_BT=1$, in this system of units time
has the dimensions of a volume, so that $1s \equiv 4 \mu m^3$). The
solution of eq.  (\ref{diffusion}) is well known: It is characterized
by dynamic correlations of the angle which diffuse along the filament
according to the law $l^2_{tw} \sim t K/a^2$. A monomer on the
filament rotates by an angle $\langle \phi^2 \rangle \sim
\sqrt{t/Ka^2}$, if $l_{tw} <L$.  When $l_{tw} >L$ the twisting motion
is sensitive to the boundary conditions on the filament; for free
boundary conditions the filament rotates freely and $\langle
\phi_{free}^2 \rangle \sim t/L a^2$.  If one end is held at a fixed
angle the other has an amplitude of rotation which saturates to
$\langle \phi_{hold}^2\rangle \sim L/K$.

Let us now calculate the writhing contributions to the spinning. For a
short filament $(L/ \kappa <1)$ oriented along the z-axis the area
enclosed by the curve ${\bf t}(s)$ is given by
\begin{math}
  A_{\Omega}=1/2 \int e_z.(\dot {\bf t} \land {\bf t}) ds.
\end{math}
The writhe of a filament is signed and averages to zero for a polymer
at thermal equilibrium. We shall consider the statistics and dynamics
of
\begin{math}
  \langle A_{\Omega}(t) A_{\Omega}(t') \rangle.
\end{math}
The linearized bending modes of the filament obey a Langevin equation
\begin{equation}
{\partial  {\bf r_{\perp}} \over \partial t} = -\kappa  {\partial^4 {\bf r_{\perp}}\over \partial s^4} + {\bf f_{\perp}} (t,s) 
\label{transdyn}
\end{equation}
with ${\bf r_{\perp}}$ the transverse fluctuations of the filaments
and ${\bf f_{\perp}}$ the Brownian noise. Using eq. (\ref{transdyn})
we find that
\begin{math}
  \langle {\bf t}_i,(q,0){\bf t}_{j}(-q,t) \rangle =
  \delta_{i,j}\exp(-\kappa q^4 t) / \kappa q^2
\end{math}
with $\{i,j\} = \{ x, y \}$. The fourth order correlation function in
$\langle A_{\Omega}(t) A_{\Omega}(t') \rangle$ can be expanded using
Wick's theorem: A short calculation shows that
 \begin{equation}
\langle A_{\Omega}^{2}\rangle = \langle W_r^2\rangle \sim L^2/ \kappa^2.
\label{wr2}
\end{equation} 
and that
\begin{equation}
 \langle (W_r(t)-W_r(0))^2\rangle \sim  l_1(t) L/ \kappa^2 \sim t^{1/4}
\label{writhe}.
\end{equation}
with $l_1(t) \sim (\kappa t)^{1/4}$ the characteristic length scale in
solutions of eq.(\ref{transdyn}). We note that the same result can be
found in a scaling approach using the following argument: The path
$\ts$ on $\s2$ is a Gaussian random walk ; On a time scale $t$ each
section of length $l_1$ of the filament re-equilibrates a random walk
with radius of gyration on $\s2$ of $r^2_{\Omega} \sim l_1/ \kappa$ and
area $A_1 \sim \pm l_1/ \kappa$.  There are $N_1= L/l_1 $ dynamically
independent contributions to the writhe variation giving $ \langle
\Delta W_r^2 \rangle \sim N_1 A_1^2 = l_1(t) L/ \kappa^2 $ as found
above. This argument shows in addition that the scaling law eq. (\ref{writhe})
is valid at short times even for filaments for which $ L>\kappa$.

What influence does this writhe fluctuation have on the global motion
of a filament. To answer this question consider two extreme,
non-physical, cases before coming back to the typical experimental
situation.  Consider, firstly, the case $K/ \kappa \gg 1$; spinning of
the end of a polymer can only occur via writhe, the twist degrees of
freedom are frozen out. The derivation of eq.  (\ref{writhe}) has
neglected rotational friction: It has been derived using a description
of only the {\it transverse} motions of the polymer. It can only be
valid when the rotational friction of a filament is so small that
spinning driven by the writhe is able to relax without build up of
torsional stress. Similar remarks have recently been made in the
propagation of tensional stress in stiff polymers \cite{us}. In this
case simple arguments allow one to deduce the validity of the
approximation made by balancing the driven motion due to the
transverse fluctuations against any additional sources of friction.
We shall now apply the same argument to the torsional motion, before
checking the results numerically. For the rotational friction to be
negligible the free rotational diffusion of a section of filament of
length $L$ must be faster faster than the driven writhing motion,
implying $\langle \phi_{free}^2 \rangle > \langle \Delta W_r^2\rangle
$ or $ t/La^2 > l_1(t) L/ \kappa^2 $.  For filaments which do not
satisfy this criterion, {\it i.e.} when $L >l_{W_r}(t)=\sqrt{l_1^3
  \kappa/a^2}$, the torsional stress has not had time to propagate
between the two ends of the filament \cite{kamien} and the law
(\ref{writhe}) does not apply.  If we observe now the end of a
filament for times shorter than the torsional equilibration time the
spinning is due to an end section of length $l_{W_r}$ and we should
substitute this effective length in eq.  (\ref{writhe}) to calculate
the end motion
\begin{math}
  \langle \Delta W_r^2 \rangle \sim {l_1^{5/2} / a \kappa ^{3/2}} \sim
  t^{5/8}.
\label{shortime}
\end{math}
It is only after the propagation of the torsional fluctuations over
the whole length of the filament that (\ref{writhe}) becomes true.  We
thus hypothesize a scaling behavior for the spinning of a filament
with $K/ \kappa \gg 1$
\begin{equation}
\langle \Psi^2 \rangle  = {L t^{1/4} \over {\kappa ^{7/4}}} {\cal Q} \left({t \kappa ^{7/3} \over a^{8/3} L^{8/3}} \right),
\label{scaling}
\end{equation} with ${\cal Q}(x)\sim x^{3/8}$ for $x$ small and  ${\cal Q}(x)\sim 1$
for $x$ large.

A second extreme case is $K=0$, as is for instance the case for many
simple numerical bead-models of worm-like chains. In this case there
is no need for the beads to rotate to follow writhe, the writhe
fluctuations are absorbed by the internal twist degree of freedom
without cost. The writhe does not lead to spinning; we do expect
however that there are strong dynamic (anti)correlations between twist
and writhe over the length of the filament. We conclude that the law
(\ref{writhe}) is valid for all times, it has no consequence however
on the rotational dynamics of a monomer on the chain that could be
detected, for instance, in a scattering experiment.

Let us now turn to the physical case, $K/ \kappa \simeq 1$. From the
expressions for $l_{W_r}$ and $l_{tw}$ we find that
$l_{tw}/l_{W_r}=\sqrt{l_1/ \kappa} \ll 1$ at short times.  On the
length scale $l_{tw}$ the twist field and the writhe of the filament
are in equilibrium and we can add the fluctuations of the twist and
writhe degrees of freedom.  Beyond $l_{tw}$ the successive section of
the polymer are dynamically decoupled: a writhe fluctuation can not be
transmitted faster than the signal transmitted by $l_{tw}$ and the
second scenario becomes valid with $l_{tw}$ playing the role of a bead
size. Beyond $l_{tw}$ the filament can writhe freely without any local
consequence on the dynamics of the chain. Thus the experimental
consequences of the writhe fluctuations are probably negligible for
times such that $l_{tw} <L$. When $l_{tw}$ reaches $L$ the amplitude
of twist fluctuations saturates, while the dynamics of the slower
bending modes continues to evolve with $\Psi$ varying as eq.
(\ref{writhe}) until $l_1(t)=L$. These additional fluctuations could
be observed experimentally.

These arguments have been checked by simulations on a discretized
bead-spring model: Take $N+1$ spherical beads of diameter $a$
connected by stiff harmonic springs.  Each bead is characterized by
its position ${\bf r}$ and a triad of orthonormal vectors. ${\bf M}=\{
{\bf b, n, t} \}$.  The vector ${\bf t}$ is an approximation to the
local tangent to the filament at ${\bf r}$. $\{ {\bf n, b} \}$ span
the normal space of the filament; ${\bf n}$ could describe the
position of some feature, such as the large groove in DNA on the
surface of the molecule. The springs are not connected to the centers
of the beads rather a bead $i$ is linked to its neighbors by
connections situated on its surface at $\{{\bf r}_i-a{\bf t}_i /2,
{\bf r}_i+a {\bf t}_i /2\}$.  The energy is
\begin{eqnarray}
E &=& B /2 \sum_i ({\bf r}_i -{\bf r}_{i+1} -a {\bf t}_i /2 + a {\bf t}_{i+1} /2)^2 \nonumber \\
  &+& \kappa/2a \sum_i (1-{\bf t}_i \cdot {\bf t}_{i+1} )\nonumber \\
  &+& K/4a \sum_i ( 1- {\bf n}_i \cdot {\bf n}_{i+1} -{\bf b}_i \cdot {\bf b}_{i+1} +{\bf t}_i \cdot {\bf t}_{i+1}) 
\label {energy}
\end{eqnarray}

The first term imposes the linear topology of the filament; when $B$
is large the curvilinear length of the filament is $L=Na$.  In the
simulations we are only interested in the limit of large $B$.  The
second term is a bending energy.  The third term is the torsional
energy. I integrate the equations of motion for the position and
orientation adding friction and thermal noise coupled to the
translational and angular velocities, effectively leading to Brownian
dynamics. Since we are particularly interested in the writhe dynamics
I use boundary conditions where the positions of the ends of the
filament are free to move but the end tangents are constrained via
external torques, so that the writhe can be calculated via Fuller's
result.

\begin{figure}[t]
  \centerline{\psfig{figure=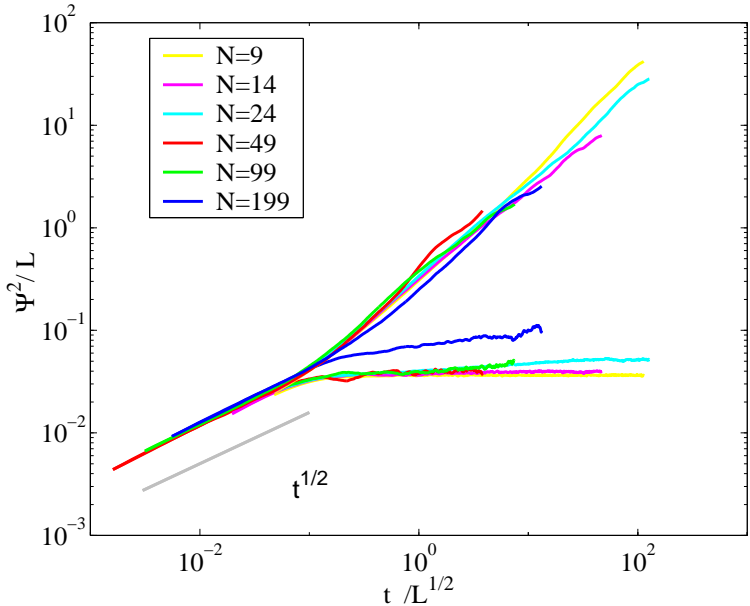,width=10.0cm}} \vskip 0.2cm
\caption{Spinning of an end as a function of time. Two different boundary
  conditions are applied at the opposite end: free spinning or held.
  At short times the boundary conditions are unimportant. The angle
  evolves as $\langle \phi^2 \rangle \sim t^{1/2}$.  For long times
  there is either saturation of the twist angle to $\langle
  \phi^2_{hold} \rangle $ (lower curves) or free diffusional rotation
  of the whole filament according to $\langle \phi^2_{free} \rangle $
  (upper curves).  $\kappa=30 a$, $L$ varies from $\kappa /3$ to $ 6.6
  \kappa$. $K/ \kappa=1$. Splitting of lower curves is due to writhe.}
\label{twistfigure}
\end{figure}

A first series of simulations were performed to study the internal
twisting dynamics of very long filaments. This simulation was
performed to confirm the intuition \cite{nelson,schurr,olddna} that
twist modes can transport torsional stress even in the presence of
large bends in the filament. This stress is carried by rapid spinning
of the filament in a slowly evolving tube, rather than the rotation of
the whole filament collectively moving against solvent friction, (like
the rotation of a rigid crankshaft). Simulations are performed for
filaments of varying length with two possible boundary conditions. The
first boundary condition is that both ends are free to spin and we
look for a crossover from the dynamics of internal modes described by
eq.  (\ref{diffusion}) to free rotation. In the second case we block
the rotation of one end and check that this leads to a saturation of
rotation angle. The scaling displayed by Fig.  (\ref{twistfigure})
shows that torsion propagates perfectly well over many persistence
lengths without hindrance from the tortuosity of the path in space.
Eq. (\ref{diffusion}) is valid for thermally bent filaments as well as
for torsional fluctuations around a linear configuration.

A second family of simulations were performed to study the writhe and
twist fluctuations of filaments both with very high torsional
constants to verify the predictions made for the writhe dynamics.
Fig(2) shows the the scaling proposed above, eq (\ref{scaling}), is
indeed seen in the case of high torsional stiffnesses.

\begin{figure}
  \centerline{\psfig{figure=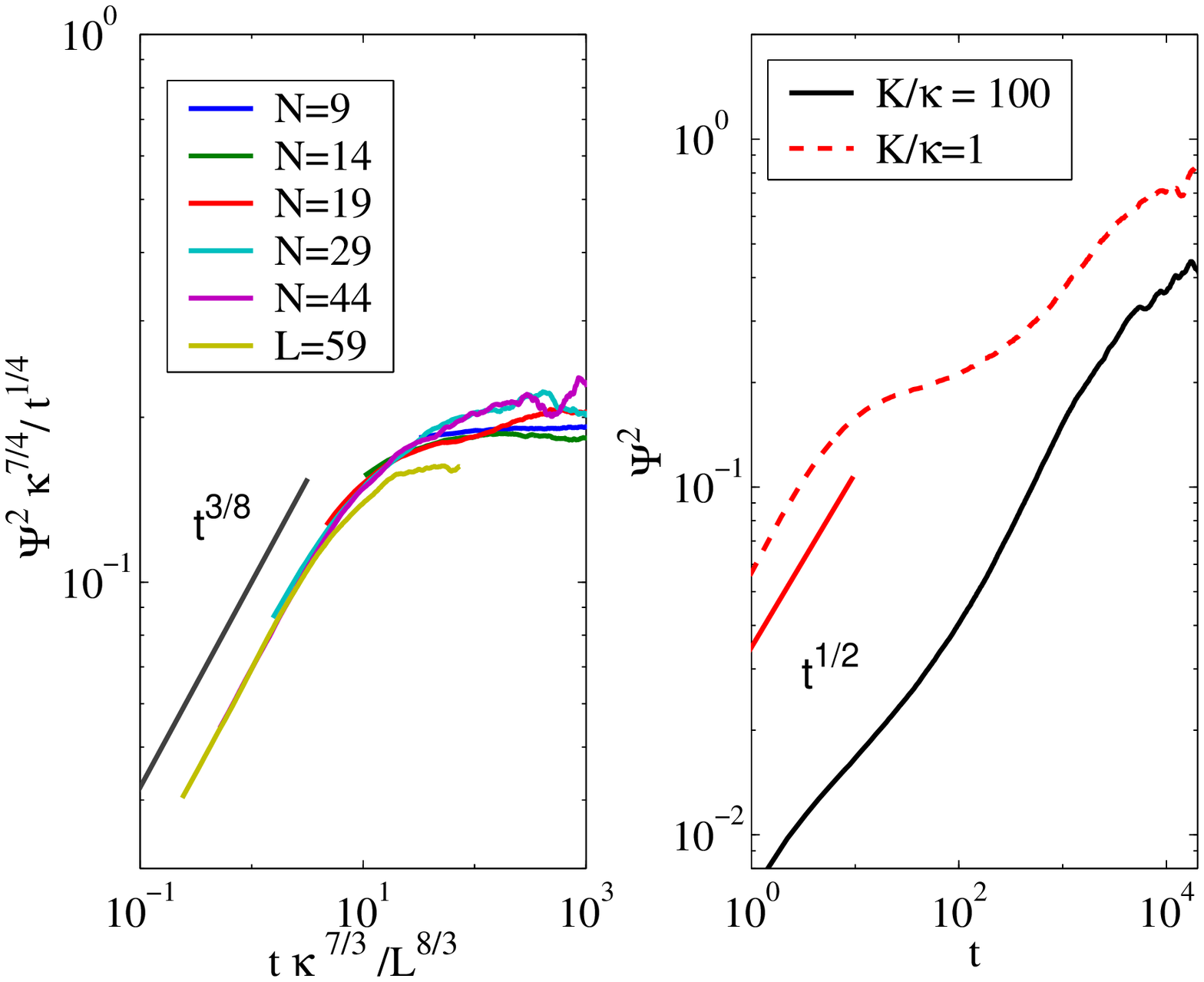,width=10.0cm}} \vskip 0.2cm
\caption{Left: Spinning of the end of a linear filament with $K/ \kappa=300$.
  Data is scaled according to eq. (5). Right: Spinning of the end of a
  semicircular section of filament for two values of $K$. In the lower
  curve all spinning is due to writhe. For the upper curve there is a
  two time behavior, with firstly twist then writhe relaxing.
  Simulations are for $L/a=\kappa/a=35$. }
\label{writhefigure}
\end{figure}

In order to calculate the writhe of a curve via Fuller's theorem we
have until now considered paths $\ts$ which are closed on $\s2$.  In
general this is most inconvenient: there are many experimental
situations where one would like to compare the spinning of filaments
oriented in an arbitrary direction. We now consider how one might
generalize the idea of writhe and spinning to filaments with arbitrary
open boundary conditions.  In general it is impossible to compare the
two rotation frames ${\bf M}(L)$ and ${\bf M}(0)$. The main ambiguity
comes from the fact that rotations are non-commutative.  However for
filaments for which $L/ \kappa \ll 1$ the non-commutative nature of
the rotations is a higher order correction and we can consider that in
going along the filament oriented along $z$ we have a rotation which
can be unambiguously decomposed into its three Cartesian coordinates
$(\Omega_x,\Omega_y,\Omega_z)$. We shall proceed by studying equations
\begin{math}
  {\partial {\bf M}(s) / \partial s} = {\bf O}(s) {\bf M}(s)
\end{math}
where ${\bf O}_{i,j}=\epsilon _{ijk}\omega^k$ with
$\omega^k=d{t}_k/ds$ for $k=\{x,y\}$.  The angular velocities
$\omega^x$ and $\omega^y$ correspond to bending the filament,
$\omega^z=0$ in the absence of twist.  We iteratively integrate this
equation.  The lowest order contribution to rotations about the z-axis
comes in second order.
\begin{equation}
  \Omega_z = {1\over 2}\int_0^L  ds ds' \  \theta(s-s')\ (\omega_s^x \omega_{s'}^y - \omega_{s'}^x \omega_s^y).
\end{equation}
Integrating by parts to transform the $\theta$-function into a
$\delta$-function gives
\begin{equation}
\Omega_z = {1\over 2} e_z.\left ( \int_0^L ( \dot {\bf t} \land {\bf t} )ds + {\bf t}(0) \land {\bf t}(L) \right)
\end{equation}
For a closed curve on $\s2$ the first term is the area enclosed by the
curve ${\bf t}(s)$ on a patch of a sphere in agreement with Fuller's
theorem. The second, boundary, term present when the curve is open
corresponds to closing the path by a geodesic from ${\bf t}(L)$ to
${\bf t}(0)$.  Thus we have a simple generalization for Fuller's
theorem valid for short chains allowing one to calculate the writhe
state of a polymer with fluctuating boundary conditions.

We remark that Fuller's theorem is closely related to Berry's phase in
experiments on polarized light transmission along bent optical fibers
\cite{haldane}. The vector $\ts$ corresponds to the local tangent to
the fiber, while the vectors, $\{{\bf n,b}\}$ transform in exactly the
same manner as the plane of polarization, via parallel transport on
$\s2$.  The problem of closing paths in $\s2$ has close analogies in
the treatment of Berry's phase in non cyclic Hamiltonians.  In quantum
and optical systems interference phenomena allow one to define the
relative phase of a system even if the evolution has not been cyclic
\cite{berry} via the Pancharatnam connection.  This convention also
closes an open trajectory again via a geodesic; in wave physics this
convention is even generally applicable and is not limited to small
paths in the appropriate projective Hilbert space.

All the above discussion has been for a filament for which the round
state is linear. Let us repeat the arguments leading to (\ref{writhe})
for short bent filaments to study the writhe fluctuations of short
curved DNA sections. The pre-existing bend considerably modifies the
arguments.  Consider a section of filament which bends an angle $\Phi$
at zero temperature due to intrinsic curvature.  At zero temperature
the bent loop has a tangent map which maps to the equator of $\s2$.
Under the map to $\s2$ a distance of $s$ in real space becomes $\Phi
s/L$.  At finite temperature there is competition between this
constant drift and the thermal agitation which leads to a Gaussian
random walk on $\s2$.  Let us now consider the dynamics in a scaling
picture: For the very shortest times only small scale structure is
changing:- the writhe dynamics reduces to the case discussed above. At
longer times the stretched path moves collectively:- each section of
length $\delta_1= \Phi l_1/ L$ on $\s2$ moves up or down by
$r_{\Omega}= \sqrt{l_1/ \kappa} $.  Counting $N_1 =L/ l_1$ dynamically
independent sections
\begin{equation}
\Delta W_r^2 \sim N_1 (\delta_1 r_{\Omega})^2 =
\Phi^2 l_1^2/ \kappa L  \sim t^{1/2}.
\end{equation}
The exponent for the writhe fluctuations has changed from 1/4 to 1/2
due to the stretching of the configuration in $\s2$. This prediction
is tested in fig. (2), where we have added a spontaneous bending
energy, in $\alpha \Sigma_i {\bf n_i.t_{i+1}}$ to curve the filament.
We now see two distinct regimes in $\Psi^2 \sim t^{1/2}$ due to the
intrinsic twist dynamics and then the writhe dynamics, the writhing
contribution is also much enhanced over its value for linear
filaments.  It would also be interesting to study the case of closed
loops, but the simple scaling argument used here becomes considerably
more complicated due to the non-local integral constraint $\int {\bf
  t}(s)\ ds =0$ coming from closure. 

This paper has studied the short time regime dominated by the bend and
twist rigidities, rather than the very long time dynamics where
crossovers to Zimm or Rouse dynamics occur.
Our arguments have
been for the simplest model of a stiff polymer, of uniform cross
section and without disorder.  The Marko-Siggia energy function
\cite{markosiggia} already has rich static behavior and one would
anticipate that the cross terms in this Hamiltonian could increase the
importance of the dynamically effects discussed here.  Similarly
disorder in the ground state is expected to dramatically modify the
dynamics, by giving a preformed writhe on $\s2$ and by modifying the
picture of simple spinning of the polymer in its tube due to new
dissipative processes; It has recently been argued that a mixture of
crankshaft and tube spinning should co-exist even in the case of weak
disorder coming from sequence fluctuations in DNA \cite{nelson}.

I would like to thank A. Ajdari, R. Everaers, F.Julicher, P. Olmstead,
C.  Wiggins and T.A. Witten for discussions.e

\end{document}